# Giant Anisotropic Magnetoresistance in Magnetic Monolayers CrPX$_3$ (X = S, Se, Te) due to symmetry breaking between the in-plane and out-of-plane crystallographic axes

W. S. Hou[1], M. Q. Dong[1], X. Zhang[2,†], Zhi-Xin Guo[1,*]

[1]*State Key Laboratory for Mechxanical Behavior of Materials, School of Materials Science and Engineering, Xi'an Jiaotong University, Xi'an, Shaanxi, 710049, China.*

[2]*Shaanxi Key Laboratory of Surface Engineering and Remanufacturing College of Mechanical and Materials Engineering Xi'an University, Xi'an 710065, China.*

†zhangxian1@outlook.com
*zxguo08@xjtu.edu.cn

**Abstract**

Anisotropic magnetoresistance (AMR) has a crucial feature for developing highly sensitive sensors and innovative memory devices. While extensively studied in bulk materials, AMR effects in these materials are typically weak. Recent advancements indicate that two-dimensional (2D) van der Waals magnetic materials possess unique magnetic properties, potentially including significant AMR characteristics. In this study, we utilize density functional theory and the Boltzmann transport equation to investigate AMR in magnetic monolayers CrPX$_3$ (X = S, Se, Te). Our findings reveal a substantially large AMR in these 2D magnetic compounds. This enhancement is attributed to magnetization (**M**)-dependent spin-orbit coupling (SOC), arising from the broken symmetry between in-plane and out-of-plane orientations. This results in significant **M**-dependent band splitting and subsequent variations in electron velocity. Additionally, we find that the **M**-dependent SOC is significantly enhanced by increasing the atomic number of the chalcogen X in CrPX$_3$, achieving an exceptional 150% AMR in CrPTe$_3$. Furthermore, our study demonstrates that AMR can be effectively modulated by applying biaxial strain, resulting in a twofold increase with a 4% strain. These findings propose a novel approach to enhancing 2D-based AMR spintronic devices, making a substantial contribution to the field.

## I. INTRODUCTION

Spin-related electromagnetic transport effects have long been a focal point of research in spintronics [1]. The electrical conductivity of magnetic materials is typically determined by their magnetic structure, resulting in the phenomenon known as magnetoresistance, with anisotropic magnetoresistance (AMR) being an intrinsic effect in magnetic materials [2-5]. Traditional AMR describes the relationship between longitudinal resistivity ($\rho_{xx}$) and the variation in magnetization direction relative to the current [6-10]. Recently, a new type of crystal-axis-dependent anisotropic magnetoresistance (CAMR) has been discovered, describing the relationship between resistivity and the variation in magnetization direction relative to the crystal axis [11-13]. Here, we collectively refer to these phenomena as AMR. AMR is generally attributed to spin-orbit coupling (SOC) and is influenced by the interaction between electronic orbital motion and spin angular momentum. AMR has significant potential applications in storage technologies and sensors, including magnetoresistive random-access memory [14,15] and giant magnetoresistive sensors [16,17].

Although a diverse range of AMR phenomena has been observed in bulk materials, most exhibit a weak AMR effect [12,18-21]. Recently, two-dimensional

(2D) magnetic materials have become a focal point in condensed matter physics research due to their intriguing properties in spintronics [22,23]. Notably, experimental studies have observed a large AMR in magnetic nanoflakes with thicknesses ranging from 15 to 46 nm [24-29], indicating a novel AMR effect induced by the reduction in spatial dimensions. However, the underlying mechanism remains to be clarified. Another urgent issue that needs to be addressed is the behavior of AMR at the 2D limit, i.e., in monolayer 2D materials.

Utilizing density functional theory (DFT) and Boltzmann transport equation (BTE) calculations, this study methodically examines the AMR effect across a range of 2D monolayer materials, specifically chromium-based phosphorus trichalcogenides ($CrPX_3$, where X represents S, Se, or Te). Although the anisotropy of 2D materials can be anticipated based on their atomic structure, the interaction between in-plane and out-of-plane symmetry breaking and AMR is nontrivial. Our investigation elucidates that in these 2D materials, the AMR effect can be significantly magnified due to robust magnetization (**M**)-dependent SOC, resulting from the spontaneous symmetry breaking that occurs between the in-plane and out-of-plane crystallographic axes. The changes in magnetization direction lead to vastly different AMR behaviors among the $CrPX_3$ monolayers, with $CrPS_3$ exhibiting a sizable negative AMR and $CrPSe_3$/$CrPTe_3$ showing large positive AMR values. This phenomenon is intricately linked to the complex p/d orbital hybridizations in $CrPX_3$. We also demonstrate that the strength of this SOC, and consequently the magnitude of the AMR effect, can be systematically increased by selecting heavier chalcogen elements for the variable X.

The large AMR value of 150% observed in CrPTe$_3$ (with M varying in the yz-plane) represents a significant departure from conventional AMR expectations. Additionally, we reveal that the AMR in 2D materials can be effectively modulated by applying biaxial strain.

## II. METHOD

Our first-principles calculations were performed using the Vienna ab initio simulation package (VASP) based on the projector-augmented wave (PAW) method [30-32]. We used the PAW pseudopotentials for the semilocal Perdew-Burke-Ernzerhof (PBE) generalized gradient approximation (GGA) [33], together with the vdW-D2 correction [34]. The Cr (3d, 4s), P (3s, 3p), S (3s, 3p), Se (4s, 4p), and Te (5s, 5p) states were treated as valence states. The convergence threshold for self-consistent-field iteration was $1.0 \times 10^{-6}$ eV, and the energy cutoff was 500 eV in all calculations [35-37]. All geometric structures were fully optimized until the convergence tolerance was achieved, with the force on each atom being less than 0.005 eV Å$^{-1}$. A Γ-centered 18×18×1 Monkhorst–Pack [38] k-mesh was used for Brillouin zone (BZ) sampling, and the vacuum layer was set to 16 Å. The GGA + U method was used to treat localized 3d orbitals, with U$_{eff}$ selected to be 4 eV for the 3d orbitals of Cr according to previous studies [39,40].

Based on the DFT calculations, the maximally localized wannier functions (MLWFs) [41–43] were constructed using the WANNIER90 code [44,45]. For the 1×1 unit cells of CrPS$_3$, CrPSe$_3$, and CrPTe$_3$, we constructed a set of 68 maximally localized MLWFs using the d orbitals of Cr, p orbitals of P, and p orbitals of X (S, Se,

Te) as the initial guess. The interpolated band structure of MLWFs is in good agreement with the results obtained from the first-principles calculations, more details see Sec. SII in Supplemental Material [51]. Based on the MLWFs, the electronic conductivity was calculated using the BTE method [46], where the chemical potential µ and temperature T dependence of electronic conductivity were obtained by

$$\sigma_{ij}(\mu, T) = e^2 \int_{-\infty}^{+\infty} d\varepsilon \left(-\frac{\partial f(\varepsilon, \mu, T)}{\partial \varepsilon}\right) \Sigma_{ij}(\varepsilon), \quad (1)$$

where $f(\varepsilon, \mu, T)$ is the Fermi-Dirac distribution function

$$f(\varepsilon, \mu, T) = \frac{1}{e^{(\varepsilon-\mu)/k_B T} + 1} \quad (2)$$

and $\Sigma_{ij}(\varepsilon)$ is the transport distribution function tensor defined as

$$\Sigma_{ij}(\varepsilon) = \frac{1}{V} \sum v_i(n,k) v_j(n,k) \tau(n,k) \delta(\varepsilon - E_{n,k}). \quad (3)$$

The sum in the above formulas is over all the energy bands (indexed by n) with all states k (including spin even if it is not explicitly denoted). $E_{n,k}$ is the energy level and $v_i(n, k)$ is the ith component of group velocity of the nth band in state k, δ is the Dirac's δ function, V = $N_k \Omega_c$ corresponds to the total volume of the system (for 2D systems, this term should eliminate the influence of the vacuum layer), and τ (n, k) is the relaxation time, which describes the average time interval between two consecutive collisions and is typically a complicated function of k and n. In our calculations, the relaxation time approximation was adopted [13,47-48], which is regarded as a constant, i.e., τ (n, k) = 1.0 × 10$^{-14}$ s. In addition, a dense k mesh of 300×300×1 was employed to perform the BZ integration for electronic conductivity calculation. According to the latest research advancements, the reliability of this method for conductivity calculation has been confirmed [13].

## III. RESULT AND DISCUSSION

The atomic structures of CrPX$_3$ (X=S, Se, Te) monolayers are depicted in Fig. 1(a). The Cr atoms exhibit a hexagonal structure analogous to graphene, with (P$_2$X$_6$)$^{4-}$-double cones fixed in place. These double cones are arranged in a triangular lattice, creating enclosed spaces for the Cr atoms. Additionally, the top and bottom chalcogen trimers exhibit a relative in-plane twist of 60 degrees. It should be noted that we primarily focus on the AMR of CrPX$_3$ monolayers in the ferromagnetic (FM) phase, which exhibits a metallic electronic structure [39]. Our DFT calculations show that the optimized lattice constants of CrPX$_3$ are 5.90 Å, 6.31 Å and 6.81 Å, and the sublayers thickness are d = 3.04 Å, 3.09 Å, 3.34 Å, for X=S, Se, and Te, respectively. These results are in good agreement with previous studies [39,40].

In the calculation of AMR, the electronic current direction is defined along the x-axis, as shown in Figs. 1(b)-1(d). The electronic conductivity $\sigma_{xx}$ (resistivity $\rho_{xx} = 1/\sigma_{xx}$) is calculated with the variation of the magnetization direction (**M**) in three planes, i.e., the xy plane (Fig. 1(b)), yz plane (Fig. 1(c)), and xz plane (Fig. 1(d)). Accordingly, the angles α, β, and γ are defined as the angles between **M** and the direction of the electronic current or crystal axis in the three planes (Figs. 1(b)-1(d)), respectively [12].

We first examine the **M**-dependent resistivity $\rho_{xx}$ of CrPX$_3$ monolayers. Figs. 2(a)-2(c) illustrate the variation of $\rho_{xx}$ with chemical potential (relative to the Fermi energy $E_F$) when **M** is along the x-axis, y-axis, and z-axis, respectively. It can be observed that $\rho_{xx}$ values with **M** along the x and y directions are nearly degenerate. However, both are significantly different from that of **M** along the z direction in certain energy regions. This result indicates that a large AMR of CrPX$_3$ monolayers can be observed with **M** varying in the xz and yz planes compared to that in the xy plane. It is noteworthy that the difference in ρ$_{xx}$ becomes increasingly pronounced as the atomic number of X (S, Se, Te) gradually increases.

We further calculate AMR values for the three CrPX$_3$ monolayers with **M** varying in the xy plane (α, red line), yz plane (β, blue line), and xz plane (γ, green line), as shown in Figs. 2(d)-2(f). The AMR is calculated using the following formula,

$$\Delta\rho_{xx} = (\rho_{xx}^{0°} - \rho_{xx}^{90°})/\rho_{xx}^{90°} \times 100\%. \qquad (4)$$

The $\rho_{xx}^{0°}$ and $\rho_{xx}^{90°}$ represent the electric resistivities when the angle between the magnetization direction and the direction of current (or crystal axis) is 0° and 90°, respectively (Figs. 1(b)-(1d)). The in-plane (xy) AMR values for all three materials are obviously smaller than those of the out-of-plane (yz, xz) ones. This feature can be understood from the perspective of the 2D structure, which has a strong correlation with the **M**-dependent SOC [49],

$$H_{SOI}^{M} = a(\vec{M} \times \vec{\nabla}U) \cdot \vec{v_c} \qquad (5)$$

where $a$ is a constant, $U$ is the electrical potential and $\vec{\nabla}U$ corresponds to effective electrical field (corresponding to the crystal axis), $\vec{M}$ represents the magnetization direction, while $\vec{v_c}$ represents the velocity of the conducting electrons influenced by an external electric field (corresponding to the applied electric field). Note that Eq. (5) is applicable to the conducting electrons with spins parallel to the magnetization direction. As shown in Fig. S2, the magnetic anisotropy energy (MAE) in the xz plane is much stronger than that of the xy plane. This result shows that the SOC is more sensitive to the out-of-plane variation of **M** due to the nature of 2D structure, leading to a more significant AMR in the xz and yz plane.

The maximum AMR value ($ARM_{max}$) for $CrPX_3$ appears with M varying in the xz and yz planes (Figs. 2(d)-2(f)). Specifically, the $ARM_{max}$ of $CrPS_3$ appears with chemical potential ($E-E_F$) equaling to 0.25 eV, reaching a maximum value of 13.8%. This value is much larger than that of typical magnetic bulk materials composed of 3d elements, where $\Delta\rho_{xx}$ is within 5% [12,18-21]. Remarkably, with increasing atomic number, the AMR of $CrPX_3$ exhibits a substantial enhancement. The maximum AMR for $CrPSe_3$ and $CrPTe_3$ reaches 46.6% and 150.7%, occurring at $E-E_F$= 0.89 eV and 0.81 eV, respectively. To provide a guidance to the experimental applications, we further present the functional relationship between AMR and charge density in section SIV of the Supplementary Materials [51]. In addition, we have calculated the electric current direction dependence of AMR, considering that the crystal structure is anisotropic in the xy plane. As shown in Fig. S4, for electric currents along both the x-axis and y-axis, the **M**-dependent electric resistance and AMR are very similar. This result implies that the AMR is not sensitive to the in-plane current direction.

Next, we explore the underlying mechanism for the extremely large AMR in $CrPX_3$ monolayers. As discussed in our previous study, AMR mainly results from the **M**-dependent energy band splitting under the effect of SOC [12,13]. Hence, we additionally calculate the energy bands of $CrPX_3$ monolayers with M along the x, y, z axes, respectively. As shown in Fig. S5, in the absence of SOC, the energy bands are independent of the magnetization direction (see Fig. S5(d)-S5(f)). Nonetheless, when the SOC effect is included, the energy bands present a magnetization-direction dependence, i.e., they are obviously nondegenerate for **M**//x (or **M**//y axis) and **M**//z axis (see Fig. 3). This result confirms that the AMR originates from the SOC-induced energy band splitting, the extent of which has a magnetization-direction dependence. Moreover, as the atomic number of X increases, the band splitting becomes more pronounced, consistent with the fact that |ARM$_{max}$| increases with the atomic number of X in CrPX$_3$ monolayers.

It should be noted that near the energy level of |ARM$_{max}$|, the band splitting is also the most pronounced (Figs. 3(a-c)), indicating a primary contribution to the AMR effect from the energy bands in the shaded regions. This is because the change of electric velocity and thus the electric resistance induced by such band splitting are the most significant. On the other hand, according to the calculation of electric conductance and the AMR definition in Eqs. (1-4), the AMR peak is not solely associated with the splitting of a single energy band along a high-symmetry k-point path. Rather, it relates to the splitting of energy bands across the entire BZ. Nonetheless, only a limited number of energy bands with significant splitting at specific k-points contribute most to the AMR, as verified by the distribution of $\Delta v$ shown in Fig. 4. This effect is most noticeable at k-points where band crossings occur.

The significant X-dependent |ARM$_{max}$| in CrPX$_3$ suggests that element X primarily contributes to the SOC and thus AMR in CrPX$_3$ monolayers. Figures 3(d)-3(f) further show the orbital projected band structures for CrPS$_3$, CrPSe$_3$, and CrPTe$_3$, respectively, where the red, blue, and green balls represent the contributions

of Cr's d, P's p, and X's p orbitals, respectively. From the figures, it is evident that the vicinity of the energy bands near the Fermi level are mainly contributed by the d orbitals of Cr and p orbitals of X. In contrast, the contribution from P's p orbitals is minimal. This observation is further supported by the orbital density of states (DOS) analysis [see Sec. SVII in Supplemental Material [51]]. Note that as the atomic number of X (S, Se, Te) increases, the contribution from the p orbitals of X atoms gradually becomes more dominant. For $CrPS_3$, the energy bands are primarily dominated by the d orbitals of Cr. Nonetheless, they are mainly attributed to the p orbitals of Te in $CrPTe_3$, and it is in between for $CrPSe_3$. As discussed above, the intrinsic AMR originates from the SOC effect, the strength of which is proportional to the atomic number of the element. The increase in dominating contributions of X's p orbitals to the energy bands qualitatively explains why $CrPTe_3$ and $CrPS_3$ have the largest and smallest AMR, respectively.

According to Eqs. (1)-(3), $\sigma_{xx}$ and thus AMR, mainly result from the magnetization-direction dependent electron velocity. To illustrate this more clearly, we further calculate the distribution of electron velocity differences in the 2D BZ induced by the variation of magnetization direction. The electron velocity can be evaluated as $\upsilon(n,k) = dE_k/\hbar dk$, where $E_k$ represents the energy, k is the wave vector, and n is band index. Hence, when we talk about the electron velocity $\upsilon(n,k)$, we mean the velocity of the n-th band with wave vector k. And one can use $\Delta\upsilon(n,k)$ to characterize the change in velocity of the n-th band at wave number k caused by a change in the direction of **M** at a fixed energy level. The electron velocity difference of **M** in the s (s = y, z) direction to that of **M** in the x direction for the n-th energy band is defined as $\Delta v_{ns}(k) = v_{nx}(k) - v_{ns}(k)$ where $v_{nx}(k)$ and $v_{ns}(k)$ are the x component of electron velocities for **M** in the x and s directions, respectively. Figure 4 shows the distribution of $\Delta v_{ny}(k)$ and $\Delta v_{nz}(k)$ at energy levels of |ARM$_{max}$|, where n is the band index for the energy bands crossing the energy level of |ARM$_{max}$|. From Figs. 4(a)-4(c), $\Delta v_{ny}(k)$ of CrPTe$_3$ is more significant than that of CrPS$_3$ and CrPSe$_3$. This result is consistent with the variation of |ARM$_{max}$| for CrPX$_3$ in the xy plane, where the largest |ARM$_{max}$| is obtained in CrPTe$_3$. Figures 4(d)-4(f) additionally show the distribution of $\Delta v_{nz}(k)$ in CrPX$_3$. A common feature is that $\Delta v_{nz}(k)$ is more sizable than $\Delta v_{ny}(k)$, showing that the AMR in the xz plane is more distinguished than that in the xy plane. Moreover, with X increasing from S to Te, $\Delta v_{nz}(k)$ becomes more and more sizable, indicating an increase of |ARM$_{max}$|. These results agree well with the M and X dependent AMRs in CrPX$_3$ shown in Fig. 2, which confirm that the AMR is mainly attributed to the variation of electron velocity induced by the change of magnetization direction under the effect of SOC.

We further explore the magnetization-angular-dependent longitudinal resistivity difference ($\delta\rho_{xx}$) of CrPX$_3$ monolayers as shown as Fig. 5, defined as $\delta\rho_{xx} = \rho_{xx} - \rho_{xx}^{min}$, where $\rho_{xx}$ is the longitudinal resistivity at different angles and $\rho_{xx}^{min}$ corresponds to the minimum resistivity. Figure 5 shows the magnetization-angular-dependent $\rho_{xx}$ at the energy levels with maximum AMR. It is observed that for all three CrPX$_3$ monolayers, $\delta\rho_{xx}$ can be well described by the formula $\delta\rho_{xx} = c_1 + c_2 cos2\gamma$, where $c_1$ and $c_2$ are constants, and $\gamma$ represents the angle between the magnetization direction and the z axis (Fig. 1(d)). This feature indicates that the AMR in 2D magnetic materials also exhibits typical biaxial symmetry [24, 25]. Note that the longitudinal resistivity also exhibits magnetic anisotropy characteristics. As shown in Fig. 5, the maximum $\rho_{xx}$ appears with **M** along the x axis in CrPS$_3$, whereas it appears with **M** along the z axis in CrPSe$_3$ and CrPTe$_3$.

Now we come to understand the change of AMR sign, with X changing from S to Se and Te in CrPX$_3$, respectively. According to the BTE formula presented in section Method, $\sigma_{xx}$ (or $1/\rho_{xx}$) directly relates to the electron velocity along the x direction. Therefore, the change in resistivity is reflected in the variation of band structure with respect to the magnetization direction [13, 50]. Let us focus on the band structure in the region with the maximum band splitting (the shaded regions in Figs. 3(a)-3(c)), which contribute most to the |ARM$_{max}$|. Figures S7(a)-S7(c) show the enlarged images of the shaded regions of band structure shown in Figs. 3(a)-3(c), which directly correspond to the |ARM$_{max}$|. As one can see, for the CrPS$_3$, the slope of the energy band is greater when the magnetization direction is along the z-axis than that along the x-axis (Fig. S7(a)). However, for the CrPSe$_3$ and CrPTe$_3$, the slope of the energy bands has a contrary behavior, as shown in Figs. S7(b) and S7(c). Since the slope of energy band directly corresponds to the electron velocity, this feature indicates $\rho_{xx}^{0°}<\rho_{xx}^{90°}$ in CrPS$_3$, while $\rho_{xx}^{0°}>\rho_{xx}^{90°}$ in CrPSe$_3$ and CrPTe$_3$. According to the definition of $\Delta\rho_{xx}$ in Eq. (4) and definition of β and γ shown in Fig. 1, one can obtain the negative and positive AMRs for CrPS$_3$ and CrPSe$_3$/CrPTe$_3$, respectively.

The change on the sign of AMR can be further understood from the point view of orbital hybridizations. In materials with significant SOC, the strength of **M**-dependent SOC can be highly anisotropic depending on the magnetization direction [13]. This anisotropy is further modulated by the hybridization between the d-orbitals of the Cr atoms and the p-orbitals of the X atoms as shown in Figs. 3 (d-f), which alters the electronic band structure and, consequently, the magnetoresistance. In $CrPS_3$, the orbital projected energy bands shows that the negative AMR appearing with E-$E_F$ around 0.25 eV is mainly attributed to the d orbital of Cr (Fig. 3(d)). While, the positive AMR in $CrPSe_3$ and $CrPTe_3$ appearing with E-$E_F$ around 0.8 eV is mainly owing to the p orbital of Se and Te atoms, respectively. As one can see from Fig. 3 and Fig. S7, in $CrPS_3$ the variation of slope of energy bands attributed by the d orbitals of Cr has a converse response to the magnetization direction, in comparation to that attributed by the p orbitals of Se and Te atoms. Such distinct energy band variation with respect to the magnetization direction can be owing to the different geometry for the hybridized p and d orbitals in $CrPX_3$. This is because the slope of energy bands in the shaded regions, where significant band splitting induced by the SOC exists, is inversely proportional to the strength of **M**-dependent SOC.

For the $CrPS_3$ with chemical potential about 0.25 eV, the $d_{yz}/d_{xz}$ orbitals contribute most to the conduction electrons as shown in Fig. S7 (d). Since the $d_{yz}/d_{xz}$ orbitals lead to larger $\vec{\nabla}U$ in z direction than that in x/y direction, one can obtain a larger the SOC with **M** in x/y direction than that in z direction (see Eq. (5)). As a result, the negative AMR is obtained due to the larger band splitting (smaller band slope and thus larger electric resistance) with **M** in x/y direction than that in z direction. As for the CrPSe$_3$ and CrPTe$_3$, since the energy bands are mainly contributed by the p$_x$/p$_y$ orbitals at the energy level of E-E$_F$=0.89 eV and 0.81eV, the larger $\vec{\nabla}U$ in x/y direction would be much stronger than that in z direction. Hence, according to Eq. (5), the larger band splitting with **M** in z direction than that in x/y direction is obtained, which leads to the positive AMR in CrPSe$_3$ and CrPTe$_3$.

Finally, we unveil the biaxial strain effect on the AMR of CrPTe$_3$ monolayers. Figure 6 shows the energy-level dependent AMR in CrPTe$_3$ around E$_F$ under tensile strain (Figs. 6(a)-6(c)) and compressive strain (Figs. 6(d)-6(f)), respectively. A common feature is that the AMR is sensitive to the applied strain. Under tensile strain, the out-of-plane AMR (with **M** in the xz and yz planes) is generally suppressed (Figs. 6(b) and 6(c)), while it is enhanced under compressive strain (Figs. 6(e) and 6(f)). Particularly, the |ARM$_{max}$| in the xz plane can be increased from 10% to 26% under a compressive strain of 4% (Fig. 6(f)). It is noteworthy that the strain effect on AMR in the xz plane is distinct from that in the yz plane, differing from the zero strain cases. Such an anisotropic response of AMR to in-plane biaxial strain can be attributed to the asymmetric geometry of CrPX$_3$ monolayers along the x and y axes. Nonetheless, for the in-plane AMR, both tensile and compressive strains tend to slightly enhance the |ARM$_{max}$| (Figs. 6(a) and 6(d)), which can be increased from 2.5% to 4.0% (5.0%) under a 3% tensile (4% compressive) strain.

Figures 7(a)-7(c) compare the electronic structure of CrPTe$_3$ without applied strain to that under a 4% tensile strain. As shown in Fig. 7(a), for the in-plane (xy)

magnetization case, the band splitting is relatively small without strain, leading to a weak AMR. The application of tensile strain promotes band splitting, enhancing the AMR effect. In contrast, for the out-of-plane (yz and xz) cases (Figs. 7(b) and 7(c)), a 4% tensile strain reduces band splitting and suppresses the AMR. Figures 7(d)-7(f) additionally illustrate that compressive strain has an opposite effect on the AMR compared to tensile strain, further highlighting the strain-induced changes in the band structure.

## IV. CONCLUSION

In summary, based on the DFT and BTE calculations for $CrPX_3$ (X = S, Se, Te) monolayers, we have unveiled the unusually large AMR in 2D magnetic materials. The large AMR arises from the strong magnetization-dependent SOC induced by the symmetry breaking between in-plane and out-of-plane directions, which induces band splitting and, consequently, electron velocity differences in different magnetization directions. The 2D geometry makes electron velocity differences particularly sensitive to the out-of-plane magnetization direction, leading to a large out-of-plane AMR. Moreover, the strength of SOC can be substantially enhanced by increasing the atomic number of X in $CrPX_3$, which gives rise to an unusually large out-of-plane AMR of 150% in $CrPTe_3$. In addition, the AMR in 2D monolayers can be effectively manipulated with biaxial strains, achieving a twofold increase in AMR by applying a 4% strain. Our results demonstrate the existence of exotic AMR properties in 2D magnetic materials, which have great potential applications in spintronic devices.

**Acknowledgments**

This work was supported by the National Natural Science Foundation of China (Nos. 12074301, 12474237), Science Fund for Distinguished Young Scholars of Shaanxi Province (No. 2024JC-JCQN-09) and the Natural Science Foundation of Shaanxi Province (Grant No. 2023-JC-QN-0768).

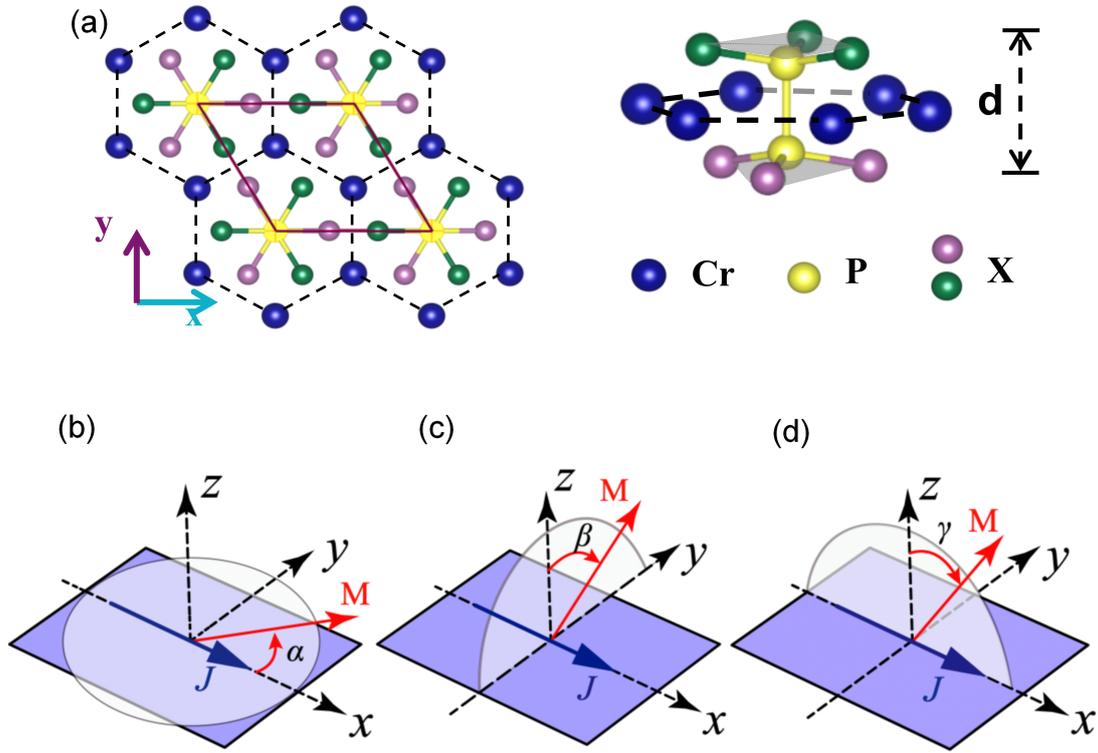

FIG. 1. (a) Schematic structure of 2D CrPX$_3$. The blue, yellow, purple, and green spheres represent chromium (Cr), phosphorus (P), and bottom and top chalcogen (X = S, Se, Te) atoms, respectively. The top and bottom chalcogen trimers have a relative in-plane twist of 60 degrees, and d represents the thickness of the sublayer. The solid red lines represent the 1×1 unit cell used for calculations. (b-d) Schematic diagram illustrating the variation of magnetization angles, with magnetization **M** varying in xy (b), yz plane (c), and xz plane (d), respectively. The blue arrow represents the electric current direction along the x-axis. The angles α, β, and γ represent the rotation angles of **M**.

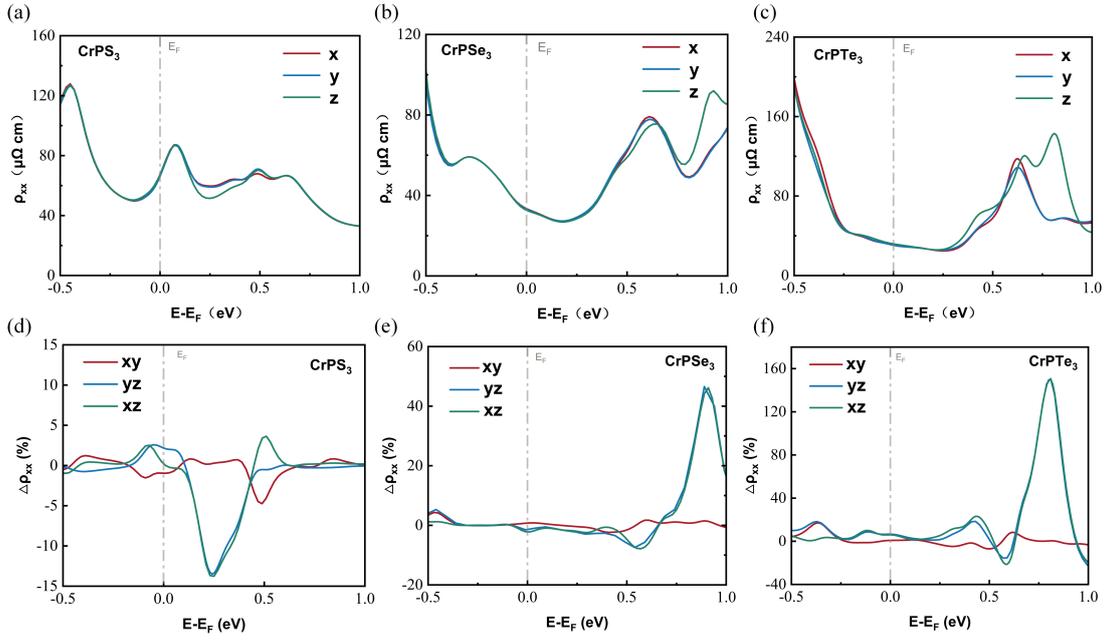

FIG. 2. (a-c) Chemical potential (E-$E_F$) dependences of resistivity for (a) $CrPS_3$, (b) $CrPSe_3$, (c) $CrPTe_3$, when the **M** is along the x-axis (red line), y-axis (blue line) and z-axis (green line), respectively. (d-f) The chemical potential dependence of $\Delta\rho_{xx} = (\rho_{xx}^{0°} - \rho_{xx}^{90°})/\rho_{xx}^{90°} \times 100\%$ for (d) $CrPS_3$, (e) $CrPSe_3$, (f) $CrPTe_3$, when the **M** varies within the xy plane (red line), yz plane (blue line), xz plane (green line), respectively.

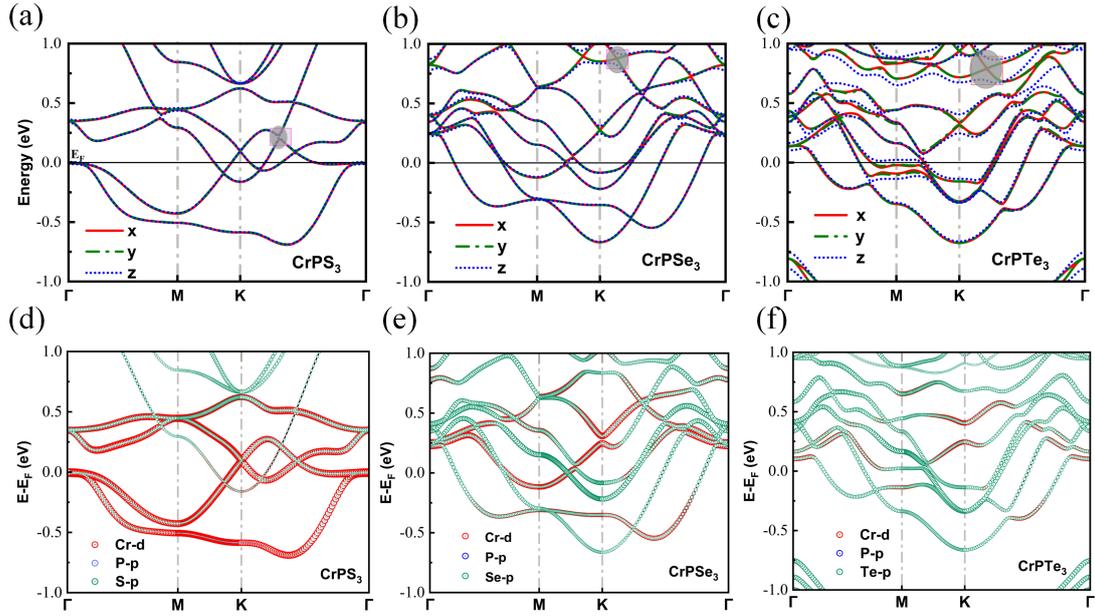

FIG. 3. (a-c) Calculated energy bands with the SOC for (a) CrPS$_3$, (b) CrPSe$_3$, and (c) CrPTe$_3$, respectively. The red solid line, green dash-dot line and blue dotted line represent the band structures of **M** along the x-axis, y-axis and z-axis, respectively. The significant split of energy bands induced by different magnetization directions is indicated by the grey circles. The red square frame is the shaded enlarged area. The fermi level is set to zero (black solid line). (d-f) Calculated energy bands with the SOC for (d) CrPS$_3$, (e) CrPSe$_3$, and (f) CrPTe$_3$ when the **M** is along the z-axis, respectively. The red, blue, and green circles represent the contributions of Cr's d orbitals, P's p orbitals, and X (S, Se, Te)'s p orbitals, respectively.

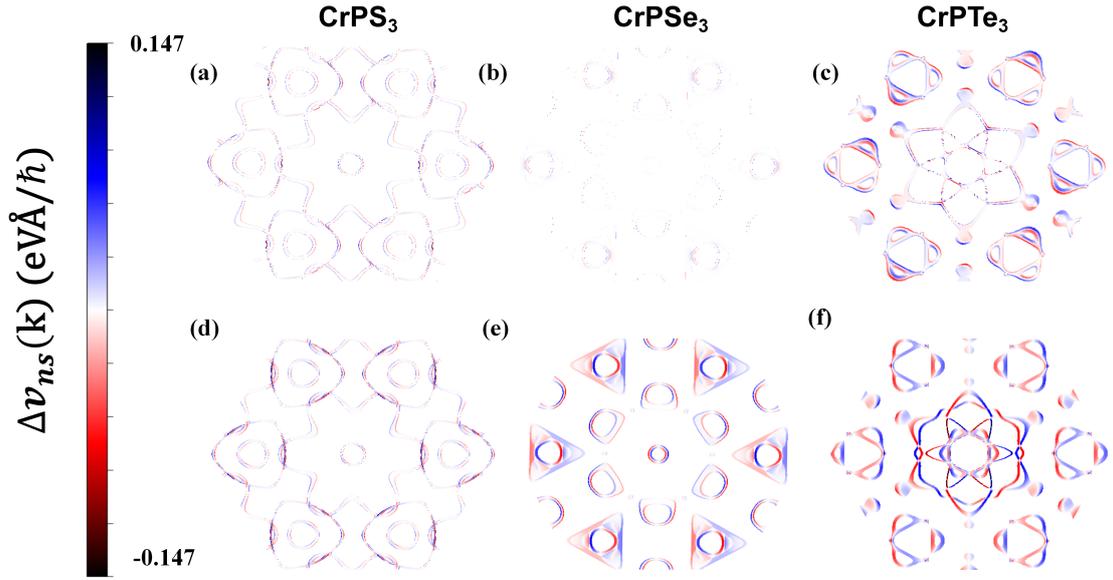

FIG. 4. The velocity differences $\Delta v_{ns}(k) = v_{nx}(k) - v_{ns}(k)$ in the electron velocity distribution at the energy level of $|ARM_{max}|$, where n is the band index for the energy bands crossing the energy level and k is the wave vector. (a-c) $\Delta v_{ny}(k) = v_{nx}(k) - v_{ny}(k)$ for (a) $CrPS_3$, (b) $CrPSe_3$, and (c) $CrPTe_3$, respectively. (d-f) $\Delta v_{nz}(k) = v_{nx}(k) - v_{nz}(k)$ for (d) $CrPS_3$, (e) $CrPSe_3$, and (f) $CrPTe_3$, respectively. $v_{nx}(k)$, $v_{ny}(k)$, $v_{nz}(k)$ are the x, y, z components of electron velocities for **M** in the x, y, z directions, respectively.

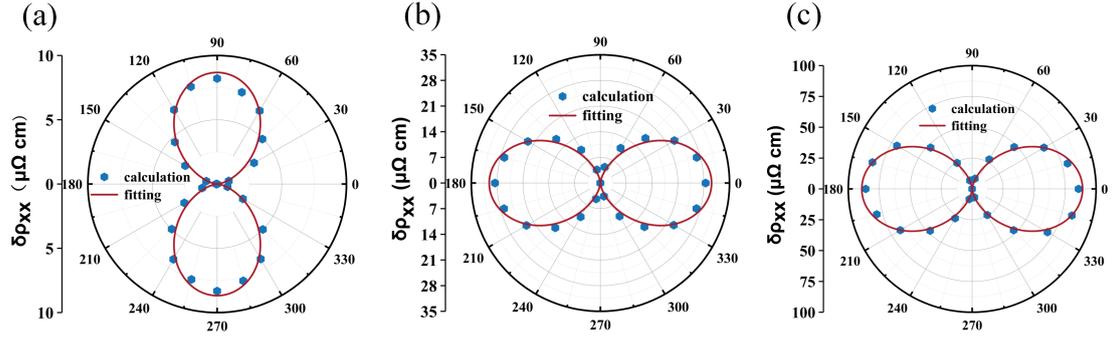

FIG. 5. Calculated longitudinal resistivity difference $\delta\rho_{xx}$ of (a) CrPS$_3$ at E-E$_F$ = 0.25 eV, (b) CrPSe$_3$ at E-E$_F$ = 0.91 eV, and (c) CrPTe$_3$ at E-E$_F$ = 0.81 eV as a function of magnetization direction $\gamma$ when **M** varies within the xz plane. The solid red lines represent results from the function fitting in use of $\delta\rho_{xx} = c_1 + c_2 cos2\gamma$.

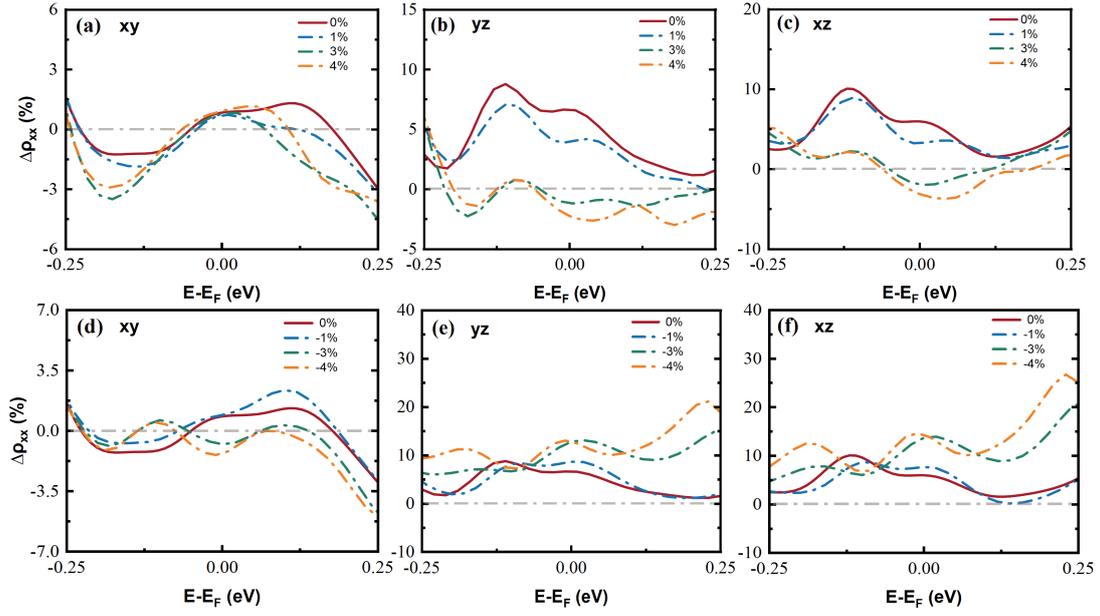

FIG. 6. Calculated $\Delta\rho_{xx}$ of CrPTe$_3$ varying with chemical potential (E-E$_F$) under different strains. The solid red line indicates that no strain is applied. (a-c) Effect of tensile strain on $\Delta\rho_{xx}$ when the magnetization direction changes in (a) xy, (b) yz, (c) xz planes, respectively. (d-f) Effect of compressive strain on $\Delta\rho_{xx}$ when the magnetization direction changes in (d) xy, (e) yz, (f) xz planes, respectively.

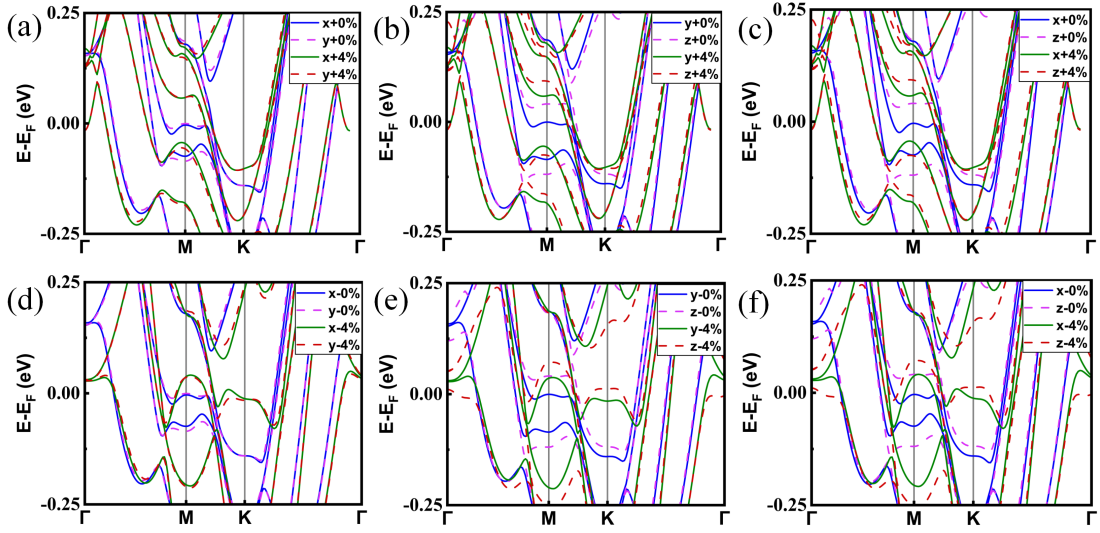

FIG. 7. (a-c) Band structures of CrPTe$_3$ under 0% and 4% tensile strains. The magnetization direction is along the x axis and y axis in (a), along the y axis and z axis in (b), and along the x axis and z axis in (c), respectively. (d-f) Band structures of CrPTe$_3$ under 0% and 4% compressive strains. The magnetization direction is along the x axis and y axis in (d), along the y axis and z axis in (e), and along the x axis and z axis in (f), respectively.